\documentclass[prb,twocolumn,showpacs,amsmath,amssymb]{revtex4}

\usepackage{epsfig,psfrag}
\usepackage{dcolumn}
\usepackage{bm}
\newcommand{\ket}[1]{\left|#1\right\rangle}
\newcommand{\braket}[2]{\langle#1|#2\rangle}
\newcommand{\brkt}[3]{\left\langle#1\right|#2\left|#3\right\rangle}

\newcommand{\refp}[1]{(\ref{#1})}
\newcommand{\Abb}[1]{Fig.~\ref{#1}}

\newcommand{\E}{electron}
\newcommand{\En}{energy}
\newcommand{\Ens}{energies}
\newcommand{\D}{density}
\newcommand{\Ds}{densities}
\newcommand{\SD}{Slater determinant}
\newcommand{\CC}{coupling constant}
\newcommand{\IA}{interaction}
\newcommand{\QP}{quantum dot}
\newcommand{\TZ}{electron number}
\newcommand{\Grdz}{ground state}
\newcommand{\YaL}{Yannouleas and Landman}

\newcommand{\HF}{\rm{HF}}
\newcommand{\QMC}{\rm QMC}
\newcommand{\GZ}{\rm GS}
\newcommand{\Be}{\rm B}
\newcommand{\exakt}{\rm exact}
\newcommand{\ges}{\rm tot}
\newcommand{\eff}{\rm eff}

\begin{document}
\title{Unrestricted Hartree-Fock for Quantum Dots}

\author{Boris Reusch}
\altaffiliation[Present address: ]{Institut f\"ur Theoretische Physik IV, 
Heinrich-Heine-Universit\"at, D-40225 D\"usseldorf, Germany}
\email{reusch@thphy.uni-duesseldorf.de}
\author{Hermann Grabert}
\affiliation{Physikalisches Institut, Albert-Ludwigs-Universit\"at, 
D-79104 Freiburg, Germany}
\date{30. October 2002}
\begin{abstract}
We present detailed results of Unrestricted Hartree-Fock (UHF)
calculations for up to eight electrons in a parabolic quantum dot.
The UHF energies are shown to provide rather accurate estimates
of the ground-state energy in the entire range of parameters
from high densities with shell model characteristics to low densities
with Wigner molecule features. To elucidate the significance of breaking 
the rotational symmetry, we compare Restricted Hartree-Fock (RHF) and UHF. 
While UHF symmetry breaking admits lower ground-state energies, misconceptions
in the interpretation of UHF densities are pointed out.
An analysis of the orbital energies shows that for very strong interaction 
the UHF Hamiltonian is equivalent to a tight-binding Hamiltonian. 
This explains why the UHF energies become nearly spin independent
in this regime while the RHF energies do not.
The UHF densities display an even-odd effect
which is related to the angular momentum of the Wigner molecule.
In a weak transversal magnetic field this even-odd effect disappears.
\end{abstract}

\maketitle

\section{Introduction}
\label{UHF}

In the present work we discuss properties, predictions, and limitations
of Hartree-Fock (HF) calculations for quantum dots.
This method has a long tradition in atomic and nuclear physics, its
application to quantum dots is therefore natural and has been
discussed in various recent papers.\cite{szafr03,sundq02,reusc01,yanno,muell96,fujit96,palac94,pfann93} 
As we will demonstrate, some of the conclusions drawn on the basis of HF 
calculations are not based on firm grounds. This is in particular the 
case, when the HF wave functions are used to describe charge 
distributions in a quantum dot. On the other hand,
Unrestricted Hartree-Fock (UHF) will be shown to give rather reliable
estimates for the ground state energies.

While quantum dots may be considered as
tunable artificial atoms, the electron density can be much smaller than in
real atoms and correlations play a more prominent role.\cite{reima02}
This is why for quantum dots the HF method has to be regarded with care.
In this work we focus on the crossover from weak to strong 
Coulomb interaction, i.e.~from higher to lower
electronic densities. This is equivalent to weakening the external
confinement potential for a given host material of the quantum dot.

The physics of this crossover can be sketched as follows:
In the case of weak interaction (high density) a one-particle picture
is valid: Electrons are filled into the energy shells of the two dimensional
isotropic harmonic oscillator. Here, the 
appropriate method is Restricted Hartree-Fock (RHF),\cite{fujit96,pfann93} 
where every orbital belongs to an energetic shell and has good 
orbital momentum. This shell filling with Hund's rule has been probed 
experimentally in small dots.\cite{taruc96}
In the case of strong interaction (low density) one can no longer
stay within this simple one-particle picture:  Wigner \cite{wigne38}
has shown that for strong correlation the ground state of the 2D electron gas
is described by localized electrons, 
representing a classical hexagonal crystal.
Accordingly, in this limit the electrons in the dot form  
a small crystal, a so-called Wigner molecule, and
the picture of energetic shells is no longer meaningful.
One has to improve the HF approximation by
passing over to UHF which means
that the space of the HF trial wave functions is extended.
The UHF Slater determinant lowers the energy by breaking the symmetry
of the problem, i.e.~spatial and spin rotational invariance.
This complicates the interpretation of the UHF solution.

For very strong \IA\ UHF is also expected to give reasonable
results because a one-particle picture of localized orbitals\cite{sundq02}
should model the Wigner molecule quite well. 
In fact, the UHF energies become nearly spin independent,
while this is not the case with RHF.
We show that the UHF Hamiltonian for strong \IA\ 
has the same spectrum as a tight-binding Hamiltonian
of a particle hopping between the sites of a Wigner molecule.
The hopping matrix elements and on-site energies can be extracted
from the UHF orbital energies.
The localization-delocalization transition has already been probed 
experimentally in larger quantum dots, \cite{zhite99}
so Wigner molecule spectrosco\-py is within reach of current technology.

An incomplete account of our results has been presented in an earlier short
communication.\cite{reusc01} Here, we discuss in detail the two-electron 
problem and present an elaborate analysis of the limit of strong interaction. 
In Sect.~\ref{QDOT} we shortly recall the model and method. 
In Sect.~\ref{QHE} we obtain explicit results for quantum-dot Helium 
that already show many features of HF solutions for higher 
electron numbers presented in Sect.~\ref{HFerg}.
In Sect.~\ref{Maguhf} we also discuss the effect of a magnetic field.

\section{Hamiltonian and Hartree-Fock approximation}
\label{QDOT}

In this work we follow the notation and method presented 
in our earlier article\cite{reusc01} for zero magnetic field.
The Hamiltonian of an isotropic parabolic \QP\ with magnetic field reads
(see e.g.~Refs.~\onlinecite{reima02,szafr03,sundq02,reima00,hiros99,ruan95,
haeus00,egger99,mikha02,reusc01,yanno,muell96,fujit96,palac94,pfann93,
chan01,ruan99,ruan00,taut93,macdo93})
\begin{equation}
H = \sum_{i=1}^N\left\{\frac{1}{2m^*}[\bm p_i + e \bm A (\bm r_i)]^2
+\frac{m^*\omega^2}{2}\bm r_i^2\right\} + \sum_{i<j}\frac{e^2/\kappa}{|\bm r_i - \bm r_j|} 
\label{hamag}
\end{equation}
where the positions (momenta) of the electrons are
denoted by $\bm r_j \; (\bm p_j)$. The effective mass is $m^*$,
and the dielectric constant is $\kappa$.
The vector potential of a homogeneous magnetic field  $\bm B$ orthogonal
to the plane of the \QP\ in symmetric gauge reads 
$\bm A (\bm r) = \frac{B}{2} (-y,x,0)$, and
the corresponding cyclotron frequency is $\omega_c= e B / m^*$.

Now we can introduce oscillator units, and describe
the system dimensionless: energies in units of $\hbar \omega_{\eff} = \hbar
\sqrt{\omega^2 + \omega_c^2/4}$ and lengths in units of
$l_0 =\sqrt{\hbar/m^*\omega_{\eff}}$.
Then the Hamiltonian takes the form
\begin{equation}
H = \sum_{i=1}^N(-\frac 1 2\triangle_i+\frac 1 2\bm r_i^2) - \frac{\omega_c}
{2\omega_{\eff}} L_z^{\ges} + \sum_{i<j}\frac{\lambda}{|\bm r_i - \bm r_j|}\;,
\label{hamag2}
\end{equation}
where we have introduced the dimensionless coupling constant
\begin{equation}
\lambda=l_0/a^*_{\Be}=e^2/\kappa l_0\hbar\omega
\end{equation}
with the effective Bohr radius $a^*_{\Be}$. For example $\lambda\!=\!2$
corresponds to $\hbar\omega\!\approx\!3$meV for a GaAs quantum dot.
The Hamiltonian (\ref{hamag2}) is formally the same as without magnetic field,
apart from an additional term proportional to the total angular momentum
which  scales with the dimensionless parameter
\cite{foot6} $\Omega_c := \omega_c /\omega_{\eff}$.
The major part of our calculations presented below is for zero magnetic field.

Regarding the HF approximation,\cite{fock30} let us recall 
the expansion of the HF orbitals in terms of the angular momentum 
eigenfunctions of the two-dimensional harmonic oscillator\cite{reusc01}
\begin{equation}
\label{hforb}
\braket{\bm{r}}{i}=\varphi_i(\bm{r}) = \sum_{\substack{n=0,\infty \\ M=-\infty,\infty}}
u_{nM}^i\braket{\bm{r}}{n M \sigma_i}\; .
\end{equation}
Here, $M$ is the angular and $n$ the radial quantum number of the
Fock-Darwin basis. 
Each orbital has its own fixed spin $\sigma_i=\pm 1/2$, 
this means there is no double occupancy of orbitals with spin up and down,
but there are different orbitals for different spins.
Thus only the $z$-component of the total spin is fixed,
$S_z^{\ges}=\sum_i \sigma_i \equiv S_z$. Furthermore, the orbitals 
(\ref{hforb}) are in general no longer eigenfunctions of the one-particle 
angular momentum (UHF). Therefore the HF Slater determinant 
is not an eigenstate of the total angular momentum $L_z^{\ges}$, 
it  breaks the symmetry of the original Hamiltonian.\cite{foot2}
Another possibility is to give each orbital $i$ 
a fixed angular momentum $M_i$. With this restriction 
one obtains RHF \cite{fujit96,pfann93} which preserves
the total angular momentum but yields higher ground-state energies.
Still another possibility is to build a \SD\ of spatially
localized orbitals for the strongly interacting case\cite{sundq02}
or of multicenter localized orbitals in high magnetic field\cite{szafr03}
and vary these orbitals to minimize the HF energy.
Our orbitals are self-consistent and are best adapted to study 
the crossover from weak to strong correlation.

In principle the orientation of the deformed symmetry-breaking
HF solution is arbitrary. This is due to the rotational
invariance of the original Hamiltonian and can be called
orientational degeneracy.
The actual UHF solution found has a special orientation
and it depends on the initial guess for the density matrix.
Often but not always the symmetry breaking is manifested
in the HF single-particle density
$n^{\HF}(\bm r) = \sum_{i=1}^N |\varphi_i(\bm{r})|^2$.
For a quantum dot in zero magnetic field,
the Hamiltonian is invariant under time reversal.
Thus we can choose real expansion coefficients $u^i_{nM}$ in \refp{hforb}.
However, then the HF one-particle density
is always symmetric to one axis.
Any arbitrary orientation can be obtained by applying 
$\exp(i\alpha L^{\ges}_z)$ to the \SD .

\section{Unrestricted Hartree-Fock for quantum-dot Helium}
\label{QHE}

In this section we present UHF energies and densities
for the two-electron quantum dot (quantum-dot Helium) at zero magnetic field
for increasing interaction strength $\lambda$.
This illustrates the basic concepts and properties of the HF approximation, 
and reveals features that are also important for higher electron numbers.
We compare with exact results obtained by diagonalization of the 
relative motion. We also compare with the RHF method, in order
to illustrate the differences to UHF.

The UHF two-electron problem has been treated previously
by \YaL .\cite{yanno} However, we find some deviations
from their results.
An extensive discussion of the RHF solution for
quantum-dot Helium at $\lambda \approx 2$ can be found in 
Ref.~\onlinecite{pfann93}.
Finally, we want to mention that the two-electron problem has
also an analytic solution in terms of a power series.\cite{taut93}

\subsection{Two-electron Slater determinant}
\label{Zweiteisla}

The Slater determinant for two electrons with $S_z=0$
is 
\begin{equation}
\label{zweiteisla}
 \Psi^{\HF} = \frac{1}{\sqrt 2} \left[
   \varphi_1(\bm r_1)\varphi_2(\bm r_2)\chi_+^1\chi_-^2 -
   \varphi_1(\bm r_2)\varphi_2(\bm r_1)\chi_+^2\chi_-^1\right] .
\end{equation}
Here we have displayed the orbital and spin parts of the
wave function explicitly, $\chi_\pm^i$
is the spin of the $i$-th electron. The state $\Psi^{\HF}$ is
generally not an eigenstate of the total spin $\bm{S}^2_{\ges}$.
In order to obtain a singlet one has to set
$\varphi_1\!=\!\varphi_2$, and thus
\begin{equation}
\Psi^{\HF} = \varphi_1(\bm r_1)\varphi_1(\bm r_2)\chi_{\rm singlet}.
\label{cshf}
\end{equation}
This restriction is also called {\it closed-shell} HF (CSHF),
because if every orbital is filled with spin up and down,
open shells are impossible.
One sees from (\ref{zweiteisla}) that the Slater determinant
violates the symmetry of the problem. For two electrons
the spin symmetry is easily restored, namely by a superposition
of two Slater determinants with spin up/down and down/up.
For the polarized case $S_z=1$, the total spin is conserved, and
the HF wave function is a product of a symmetric spin function
and an antisymmetric orbital function.

\subsection{Different HF approximations}
\label{diffhf}
\begin{figure}[t]
{\epsfig{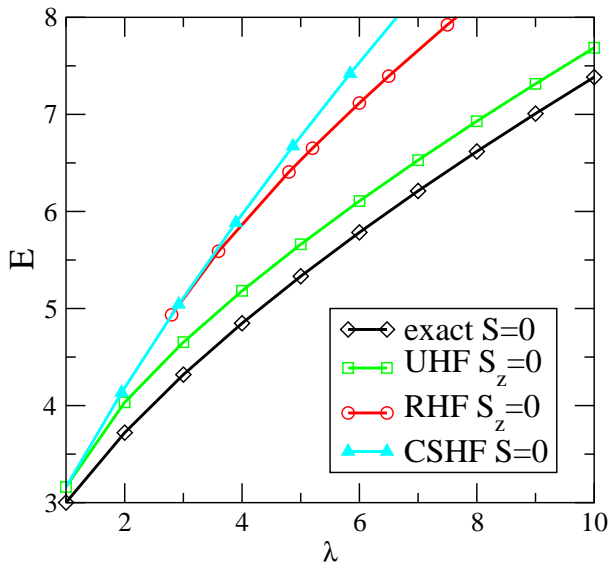}}
\caption{Comparison of different HF energies for quantum-dot Helium 
with the exact ground state energy vs.~the coupling constant $\lambda$.}
\label{s_z0}
\end{figure}
We now compare the energies of different HF approximations
with the results of an exact diagonalization.\cite{reusc01}
First we consider the case $S_z=0$. The most general ansatz for
the HF orbitals is the UHF state (\ref{hforb}), a spin dependent expansion with
arbitrary angular momentum. Less general is the RHF ansatz,
where angular momentum is preserved. And still less general is
CSHF \refp{cshf}, when we force the two electrons to occupy two identical
(rotationally symmetric) orbitals. 
In \Abb{s_z0} one can clearly see
the importance of breaking the symmetry to obtain
lower HF energies. Up to $\lambda\approx 1$ all 
three methods give nearly the same result. Up to $\lambda\approx
3$ the closed-shell energy is equal to the RHF energy. In other
words: From this point on the two RHF orbitals are no longer
identical. As expected the UHF energy is lowest.

In \Abb{sintridiff} we show the differences of the RHF and UHF energies
from the energy of the exact ground state which is the singlet.
For $S_z=1$ one needs two different orbitals, there is no CSHF. The UHF
method gives lower energies than RHF, but the gain in energy is not as big
as in the unpolarized case.
Interestingly, the UHF energies become spin independent with
increasing $\lambda$: they agree within about $0.3\%$, 
the $S_z=1$ state is somewhat lower than the $S_z=0$ state.
\begin{figure}
\epsfig{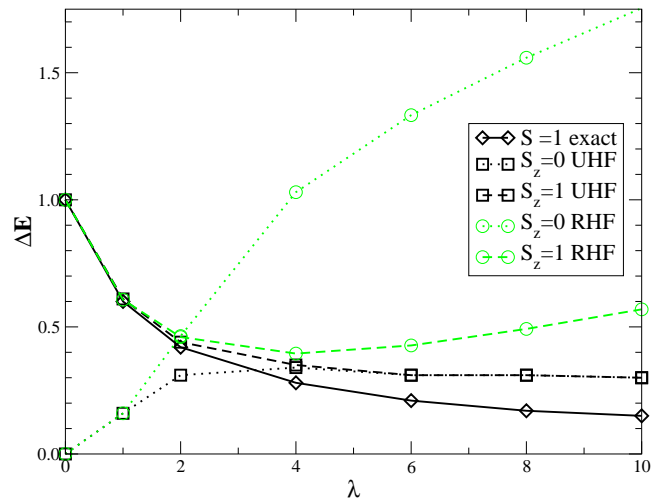}
\caption{Absolute energy differences with respect to the exact 
$S=0$ ground state  $\Delta E = E_S - E_{\GZ}^{\exakt}$.
Above $\lambda \approx 4$ the two UHF energies are nearly the same.}
\label{sintridiff}
\end{figure}
The exact energies merge more slowly: for $\lambda=20$ the energy 
difference between singlet and triplet is still about $1\%$.
Note that the RHF energies fail to become spin independent 
for large $\lambda$, as can be seen from \Abb{sintridiff}.
Of course, one expects spin independent energies in the classical
limit of localized electrons without overlap.

\subsection{UHF one-particle densities}
\label{HFhedich}

Now  we want to have a closer look at the one-particle density
which is just the sum of the densities of the two orbitals,
$n^{\HF}(\bm r) = |\varphi_1(\bm{r})|^2 + |\varphi_2(\bm{r})|^2$.
\begin{figure}
\psfrag{a}{(a)}
\psfrag{b}{(b)}
\psfrag{c}{(c)}
\psfrag{d}{(d)}
\epsfig{file=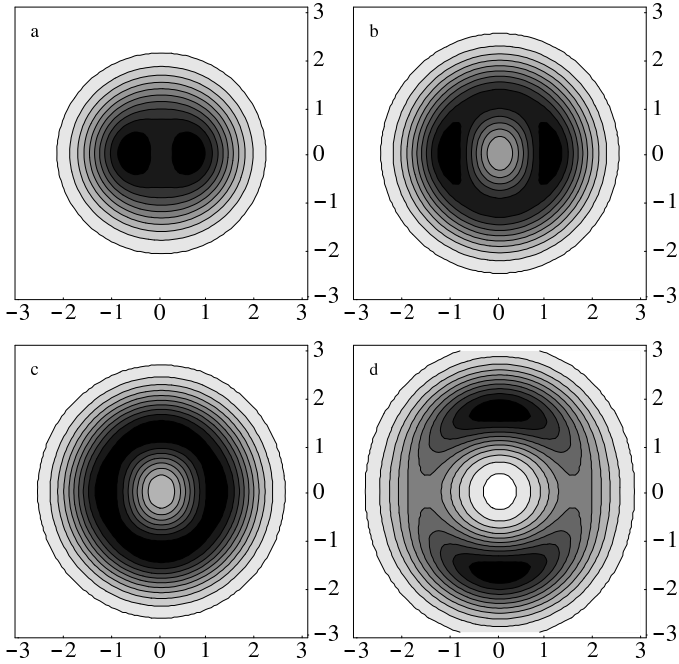, width=9cm}
\caption{Shadowed contour plots of the UHF one-particle densities $n^{\HF}$
for $N=2$, $S_z=0$. One contour corresponds to $1/10$ of the maximal density.
(a) $\lambda=2$, (b) $\lambda=6$, (c) $\lambda=8$, (d) $\lambda=20$.}
\label{n2s0dich}
\end{figure}

\begin{figure}
\psfrag{a}{(a)}
\psfrag{b}{(b)}
\psfrag{c}{(c)}
\psfrag{d}{(d)}
\epsfig{file=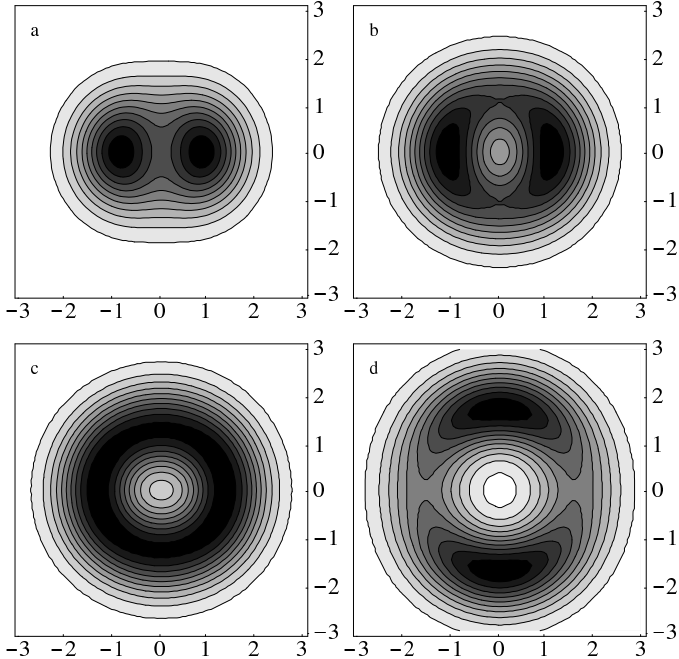, width=9cm}
\caption{UHF one-particle densities for $N=2$, 
$S_z=1$. (a) $\lambda=2$, (b) $\lambda=6$, (c) $\lambda=8$, (d) $\lambda=20$.}
\label{n2s1dich}
\end{figure}

In Figs. \ref{n2s0dich} and \ref{n2s1dich}
we show this density for different values
of the coupling parameter $\lambda$. Already for a relatively
small $\lambda$ we detect two azimuthal maxima. The density is
strongly anisotropic which is due to the symmetry
breaking.
In the case of $S_z=1$ the two maxima are
more distinct as a consequence of the Pauli principle: spin-polarized 
electrons are more strongly correlated. However, the direct
interpretation of the two dips as localized electrons is questionable.
With increasing $\lambda$ the azimuthal modulation first decreases, but
for $\lambda\agt 8$ ( $\lambda\agt 10$ for $S_z=1$) it
increases again. For very high $\lambda$ the densities
become almost spin independent.
A closer view reveals that the azimuthal maxima
are more distinct for the case $S_z=0$. This arises
from the exchange term in the HF energy: it lowers the
energy for strong interaction and overlapping spin-polarized
orbitals.

While the azimuthal modulation is an artifact of the HF method,
the densities display correctly a minimum in the center which gets deeper 
with stronger interaction. Also, the maxima are in very good
agreement with the classical positions $r_a =\sqrt[3]{\lambda /4}$
(see Appendix \ref{Klako}). 

\subsection{UHF orbitals and orbital energies}
\label{HForb}

In order to understand the form of the UHF densities it is
useful to have a closer look at the UHF orbitals.
For $S_z=0$ we find two orbitals that are exactly complex
conjugate, $\varphi_1 = \varphi_2^*$.
This can be seen by studying the expansion coefficients 
$u_{nM}$ in Eq.~(\ref{hforb}) and means that the \SD\ is symmetric 
under time reversal.
\begin{figure}
\psfrag{a}{(a)}
\psfrag{b}{(b)}
\psfrag{c}{(c)}
\psfrag{d}{(d)}
\epsfig{file=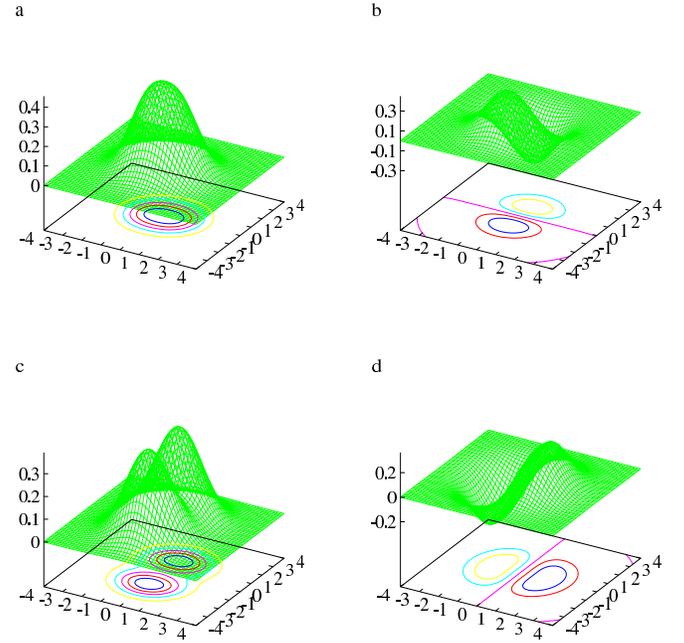, width=9cm}
\caption{Pairs of real UHF orbitals for $N=2$, $S_z=1$.
(a), (b) at $\lambda=2$, (c), (d) at $\lambda=10$.}
\label{orbn2s1}
\end{figure}

For $S_z=1$ the two orbitals depicted in \Abb{orbn2s1}
are always different and can be chosen real. 
For $\lambda=2$ one can still interpret the orbitals in the
energy shell picture of RHF: the first orbital is
(approximately) round, S-like, and the second one is dumbbell formed,
P-like.
\cite{foot2} 

For very high $\lambda \agt 14$ there is a simple relation between
the orbitals for the two spin polarizations: for $S_z=1$ we may choose
both orbitals real and then we find
\begin{equation}
\label{n2orbtra}
\varphi_{1/2}^{S=0} \approx \frac{1}{\sqrt 2} (\varphi_{1}^{S=1} \pm i\varphi_{2}^{S=1})\; .
\end{equation}
In this fashion, we see that $\varphi_{1/2}^{S=0}$ are complex
conjugate and approximately orthonormal.

To shed more light on this behavior we consider also
the orbital energies. We start with the HF Hamiltonian in the HF
basis for $S_z=1$
\begin{equation}
\label{n2s1hfha}
\left( \begin{array}{cc} \varepsilon_1 & 0 \\
			      0 & \varepsilon_2 \\ \end{array} \right)
= \left( \begin{array}{cc} h_{1 1} + w_{1212}  & 0 \\
			      0 &  h_{2 2} + w_{1212}\\ \end{array} \right)
 \; .
\end{equation}
Here, we use the notation $h_{ij}=\brkt{i}{h}{j}$ and
$w_{ijkl}=(i j|w|k l)$ for matrix elements in the HF basis (see 
Ref.~\onlinecite{reusc01}). When we apply the unitary transform 
\refp{n2orbtra}
\begin{equation}
\label{n2s1hfhatra}
\frac{1}{2}
\left( \begin{array}{rr} 1 & i \\ 1 & -i\\ \end{array} \right)
\left( \begin{array}{cc} \varepsilon_1 & 0 \\
			      0 & \varepsilon_2 \\ \end{array} \right)
\left( \begin{array}{rr} 1 & 1 \\ -i & i\\ \end{array} \right)
= \left( \begin{array}{rr} U & -t \\ -t & U\\ \end{array} \right)
= H_2 \; 
\end{equation}
we obtain a two-state Hamiltonian $H_2$, with on-site energy
$U\!=\!(h_{1 1}+h_{2 2} + 2w_{1212})/2$ and tunnel splitting
$t\!=\!(h_{2 2}-h_{1 1})/2$. Thereby, we have mapped the
HF Hamiltonian on a lattice problem.It is intuitive that for strong \IA\
the two electrons localize, and thus a tight-binding approach 
should become physically correct. 
This is also the case for larger electron number as discussed below.

\subsection{UHF two-particle densities}
\label{HFzweiteidich}

\begin{figure}
\psfrag{a}{(a)}
\psfrag{b}{(b)}
\psfrag{c}{(c)}
\psfrag{d}{(d)}
\psfrag{x}{$\!\otimes$}
\epsfig{file=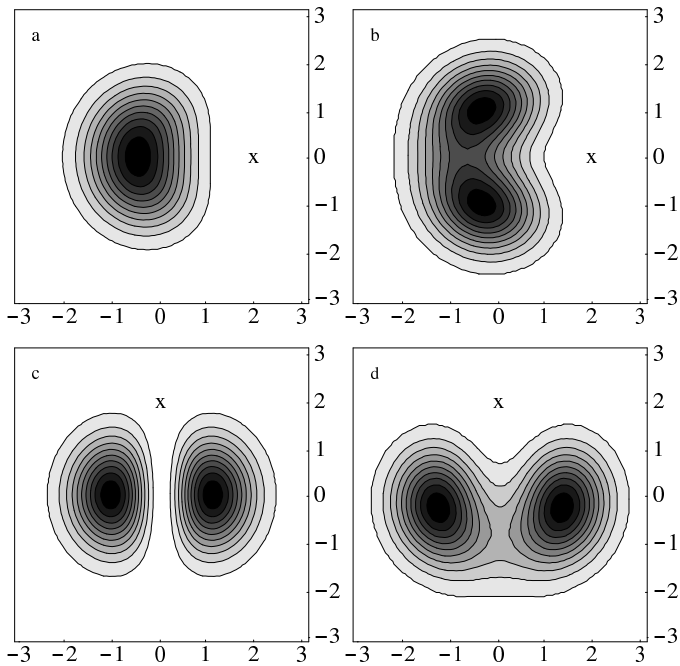, width=9cm} 
\caption{UHF Conditional probability density $n^{\HF}(\bm{x}|\bm{y})$
for $N=2$, $S_z=1$. In the upper row  $\bm{y}=(2,0)$ (x) 
(a) $\lambda=2$, (b) $\lambda=6$. Lower row  $\bm{y}=(0,2)$ (x)
(c) $\lambda=2$, (d) $\lambda=10$.}
\label{n2cpd}
\end{figure}

Next we examine the conditional probability density (CPD)
for finding one electron at $\bm{x}$, under the condition that 
another electron is at $\bm{y}$. For quantum-dot Helium and $S_z=0$ 
the CPD reads
\begin{equation}
\label{cpdn2}
 n^{\HF}(\bm x|\bm{y}) =  \frac{|\varphi_1(\bm{x})|^2
 |\varphi_2(\bm{y})|^2 + |\varphi_1(\bm{y})|^2
 |\varphi_2(\bm{x})|^2}{n^{\HF}(\bm y)} \;.
\end{equation}
Now, since we found complex conjugate orbitals, $\varphi_1 = \varphi_2^*$, 
we have $n^{\HF}(\bm x|\bm{y}) = n^{\HF}(\bm x)$, i.e.~the
conditional probability density is independent of the condition.
This is not really astonishing, because within the HF method two
electrons are only correlated by the exchange term, which
vanishes here.
\cite{foot2a}

For $S_z=1$ the orbitals are different from each other and the CPD
is given by
\begin{eqnarray}
\label{cpdn2s1}
 n^{\HF}(\bm x|\bm{y}) =  \{ |\varphi_1(\bm{x})|^2
 |\varphi_2(\bm{y})|^2 + |\varphi_1(\bm{y})|^2
 |\varphi_2(\bm{x})|^2 \nonumber \\ 
 - 2{\rm Re}[\varphi_1^*(\bm{x})\varphi_2(\bm{x})\varphi_1(\bm{y})
 \varphi_2^*(\bm{y})] \} / n^{\HF}(\bm y) \;.\;
\end{eqnarray}

In \Abb{n2cpd} we show contour plots of  UHF CPDs for
different coupling constants and given positions $\bm{y}$. In
the upper row, for $\bm{y}=(2,0)$, we find for small
$\lambda=2$ a suggestive result: the density has a single maximum
at a distinct distance from the fixed coordinate $\bm{y}$. With increasing
$\lambda$, however, we obtain two maxima, which develop more and
more and are not at all located at the classical position.

The situation is likewise irritating when one chooses $
\bm{y}=(0,2)$  as fixed coordinate (lower row). While the exact CPD is
rotationally symmetric when both $\bm{x}$ and $\bm{y}$ are rotated, 
the UHF CPD does not respect this symmetry. The reason for this lies
in the symmetry breaking which cannot completely account for correlations. 
The UHF Slater determinant is
deformed and derived quantities do not necessarily have a direct physical
meaning, -- except for the UHF energy which is a true upper bound for
the exact energy. 

\section{Unrestricted Hartree-Fock for higher electron numbers}
\label{HFerg}

In this section we show further results of UHF calculations, 
namely \Ens\ and \Ds\ for up to eight \E s ($B=0$).
Many effects are similar to what we have already seen
for two \E s, for example the errors of the UHF \Ens\ and their
spin dependence. An interesting phenomenon shown by the
UHF \Ds\ is the even-odd effect discussed below.

\subsection{UHF \Ens}
\label{HFenergien}

\begin{figure}
\epsfig{file=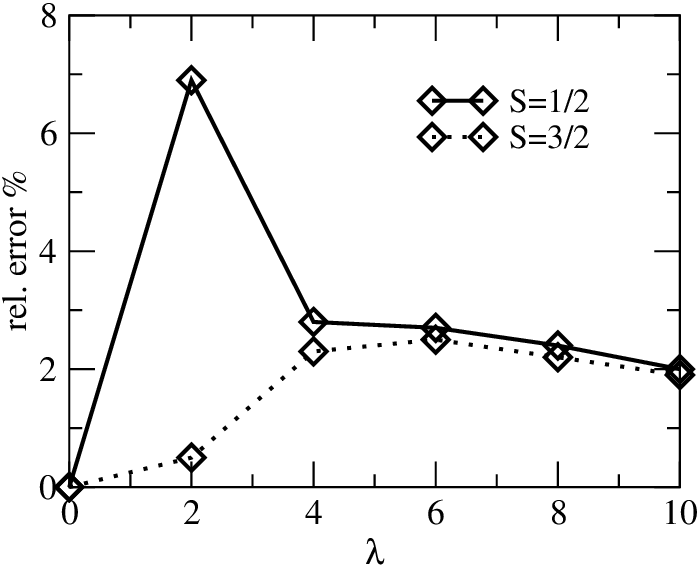, width=8.5cm}
\caption{Relative error of the UHF \En\ $(E_S^{\HF}-E_S^{\QMC})/E_S^{\QMC}$ 
for $N=3$  vs.~\CC\ $\lambda$.}
\label{rel}
\end{figure}
For $N>2$ we compare the UHF \Ens\ with results of a 
Quantum Monte Carlo (QMC) simulation by Egger et al.\cite{egger99}
These results were obtained for a very low temperature
$T\!=\!0.1\hbar\omega/k_{\Be}$. The QMC \Ens\
are always below the HF \Ens\ and can therefore
be considered as effective zero temperature reference points.

For $N=3$ QMC, a semiclassical analysis, \cite{haeus00}
as well as an exact diagonalization study \cite{mikha02} 
predict a transition from the $S=1/2$ \Grdz\ 
in the weakly interacting case to a $S=3/2$ \Grdz\ for $\lambda\agt 4$. 
Within UHF this transition occurs already near $\lambda = 2$.
In \Abb{rel} one can see that the relative error for $S_z=3/2$ 
is small, less than 3\% .
In the non-polarized case the error is higher, about 
~7\% for $\lambda\agt 2$. With increasing $N$ and
$\lambda$ the relative error becomes smaller because the absolute
\Ens\ are higher.

In \Abb{diff} we show the absolute \En\ differences
from the QMC \Grdz\ for eight \E s. 
For intermediate values of $\lambda$ the UHF \Ens\ become
already nearly spin independent, whereas the QMC \Ens\ approach 
this semiclassical behavior more slowly. 
For stronger \IA\ the HF \Grdz\ is always spin-polarized.
Thus the UHF method can not resolve the correct spin ordering
of the \Ens . 

\begin{figure}
\epsfig{file=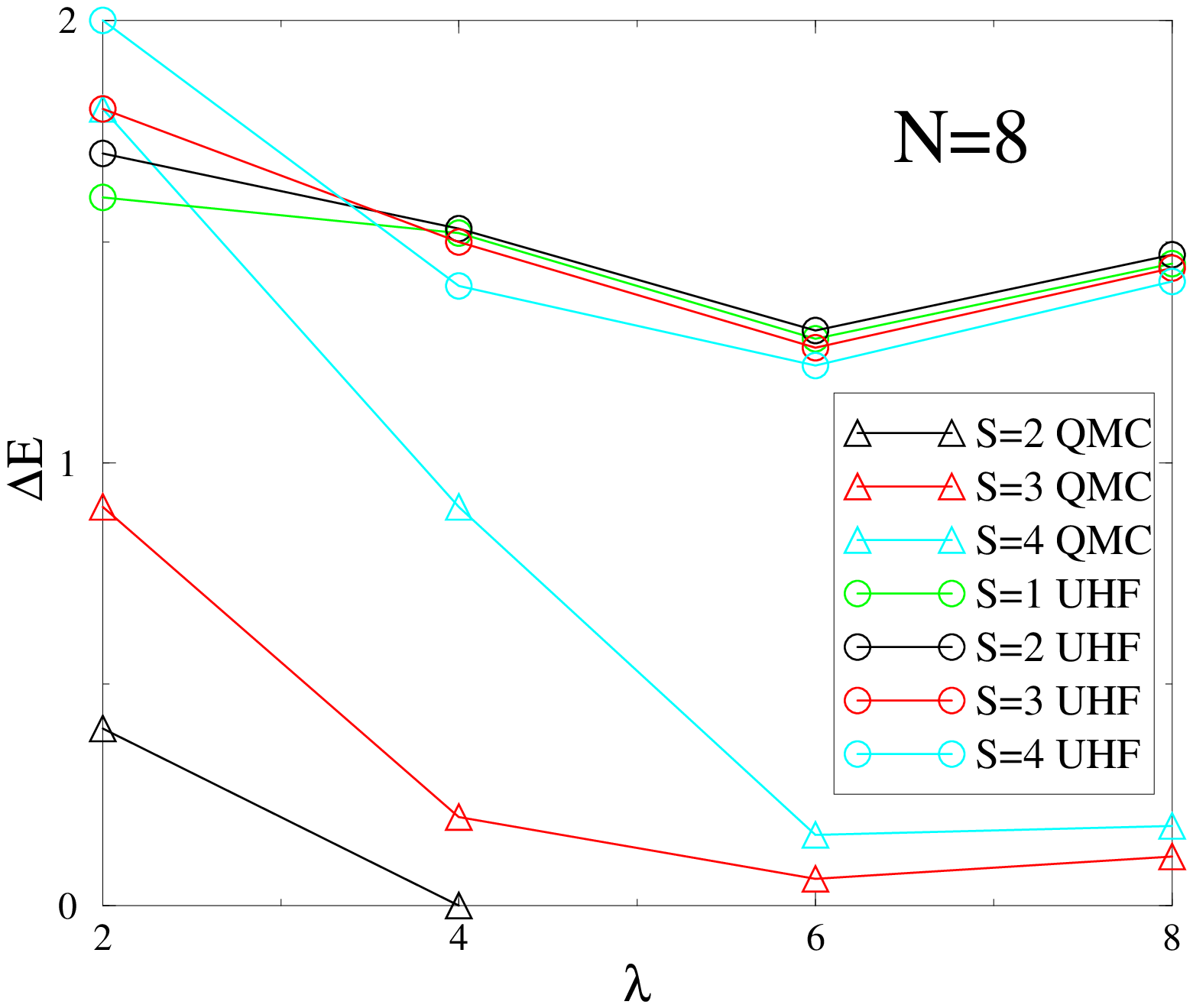, width=8.5cm}
\caption{Absolute energy differences from the QMC \Grdz , 
$\Delta E = E_S-E_{\GZ}^{\QMC}$ for eight \E s
and various spins vs. \CC\ $\lambda$.}
\label{diff}
\end{figure}

For $N\! =\! 8$ the QMC method predicts a crossover of the total spin from
$S\! =\! 1$ to $S\! =\! 2$ near $\lambda = 4$.
The UHF method finds a polarized \Grdz\ with $S\! =\! 4$ for 
$\lambda\agt 4$. There, however, the energy differences 
for different spins are already quite small.

One can conclude that the UHF \SD\ with fixed spin structure 
gives a rather poor description of the total many-electron wave function.
Essentially, UHF renders the properties of the spin-polarized solution
for larger $\lambda$. This can also be seen in the UHF \Ds , 
which become spin independent for larger \IA\ (see below).
Finally, we briefly mention the RHF results: there for large $\lambda$
the HF \Ens\ do not become spin independent, 
but the energies for lower spins are considerably higher. 
For large $\lambda$ RHF gives a poor estimate of the ground state
energy.

\subsection{HF densities: Even-odd effect}
\label{EO}

In this subsection we consider the UHF densities for
higher \TZ s. 
We first show in \Abb{evenodd} the \Ds\ 
for rather strong \CC\  $\lambda=6$, various \E\ numbers $N$
and $S_z=N/2$. Above this \IA\ strength the 
UHF \Ds\ are essentially the same for all $S_z$ (except for $N=2$, 
see above) and do not change qualitatively with
increasing $\lambda$.
\begin{figure}
\psfrag{a}{(a)}
\psfrag{b}{(b)}
\psfrag{c}{(c)}
\psfrag{d}{(d)}
\epsfig{file=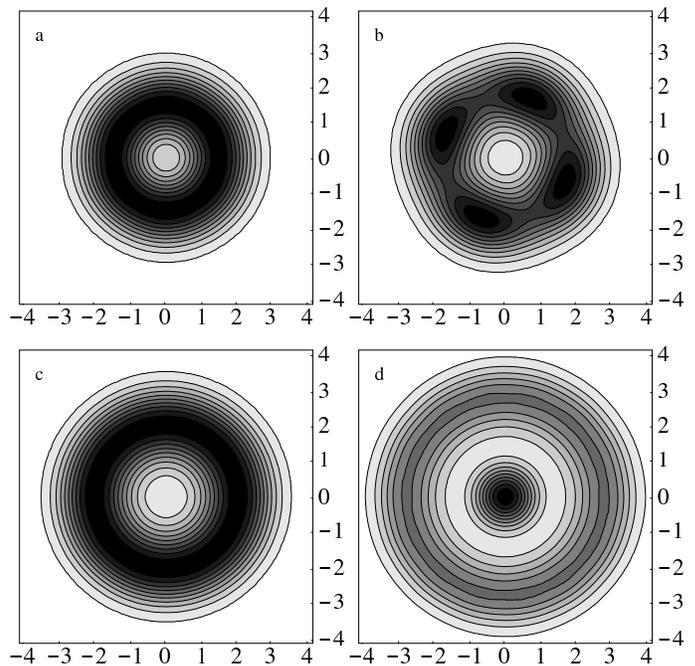, width=9cm}
\caption{Even-odd effect of the UHF one-particle \Ds\  $n^{\HF}$ for 
$\lambda=6$, different \TZ s $N$ and polarized spin $S_z=N/2$. 
(a) $N=3$, (b) $N=4$, (c) $N=5$, (d) $N=8$.}
\label{evenodd}
\end{figure}

Surprisingly, only for some $N$ does one obtain
a molecule-like structure, i.e.~an azimuthal modulation
as seen for two \E s.
For three and five \E s the \D\ is apparently rotationally symmetric
and also for eight \E s, where we have a pronounced maximum in the
center. The expected molecule-like structure shows up only for $N=2$ and 4.
Thus, when we consider also $N=6,7$ (see below)
we recognize that azimuthal maxima occur only
for an even number of \E s per spatial shell.
In stating this we want to emphasize, that all the densities shown
belong to symmetry breaking, deformed \SD s.

This even-odd effect is also surprising, because 
UHF calculations for \QP s in a strong magnetic field 
\cite{muell96} found molecule-like \Ds\ for all \TZ s,
and frequently a magnetic field leads to similar effects as a
stronger interaction.
We also have performed calculations with a magnetic field
that reproduce the \Ds \ of Ref.~\onlinecite{muell96} 
and show that the molecule-like structure disappears for odd $N$
for vanishing field.
\cite{foot3} 

A physical explanation of the even-odd effect combines 
the geometry of the classical system with the symmetry of quantum 
mechanics. \cite{ruan95} Consider the {\it exact} spin-polarized $N$-electron 
wave function $\Psi_N$ for the Wigner molecule case. 
Due to the strong Coulomb repulsion, the electrons move on an $N$-fold 
equilateral polygon (for $N<6$; for $N=6$ one electron enters the center 
of the dot). A rotation by $2\pi/N$ therefore corresponds
to a cyclic permutation
\begin{equation}
\label{geosym}
\exp \left\{\frac{2\pi i}{N} \; L^{\ges}_z \right\} \; \Psi_N = (-1)^{N-1} \; \Psi_N\;,
\end{equation}
where we have used that a cyclic permutation of an even (odd) number
of electrons is odd (even). From Eq.~(\ref{geosym}) the allowed total angular
momenta of the Wigner molecule can be easily read off: for an odd 
number of electrons the minimal angular momentum is zero, whereas it is
nonzero and degenerate for an even electron number, e.g.~$M^{\ges}
 = \pm 2$ for $N=4$.
Hence, the UHF wave functions for $N=2,4,7$
can be interpreted as standing waves, i.e.~superpositions of opposite
angular momentum states. For odd numbers of \E s in a spatial shell
there is no angular momentum degeneracy and therefore no standing wave
and no modulation in the densities.
With a similar argument Hirose and Wingreen \cite{hiros99} 
explain the charge-density-waves which they found for {\it odd} 
number of electrons in the {\it weakly} interacting regime from
density functional calculations.

Equation \refp{geosym} does not hold anymore
when the spins are not polarized, because the total
wave function is not a product of spin and orbital wave functions.
However, within UHF we do not fix the exact spin but only
subspaces with fixed $S_z$. For $S_z < N/2$ and strong interaction 
the UHF solution mainly renders the properties
of the spin-polarized solution, since the energies and \Ds\
are essentially the same for $\lambda\agt 6$.
The even-odd effect is thus {\it not} a physical effect
but an {\it artifact} of the UHF symmetry breaking.
Therefore great caution must be taken when interpreting the UHF \Ds .
In particular, the exact onset of Wigner crystallization cannot be determined
reliably from UHF calculations.

\subsection{Closer look at three \E s}
\label{HFn3}

As we have just discussed, for
three \E s with strong \IA\ we do not find the naively expected density
with three maxima but a nearly round \D .
When we plot the \D\ of \Abb{evenodd}(a) with more contour lines (not shown)
a tiny sixfold modulation of the \D\ is discernible.
This can be understood by going back to Eq.~\refp{geosym}:
after $M^{\ges} = 0 $ the next allowed total angular momentum values are
$M^{\ges} = \pm 3$, which give rise to a standing wave with six maxima.
This becomes also clear from the \Ds\ of the single
orbitals building the UHF single-particle \D .
\begin{figure*}[ht!]
\psfrag{a}{(a)}
\psfrag{b}{(b)}
\psfrag{c}{(c)}
\psfrag{d}{(d)}
\psfrag{e}{(e)}
\psfrag{f}{(f)}
\centering \epsfig{file=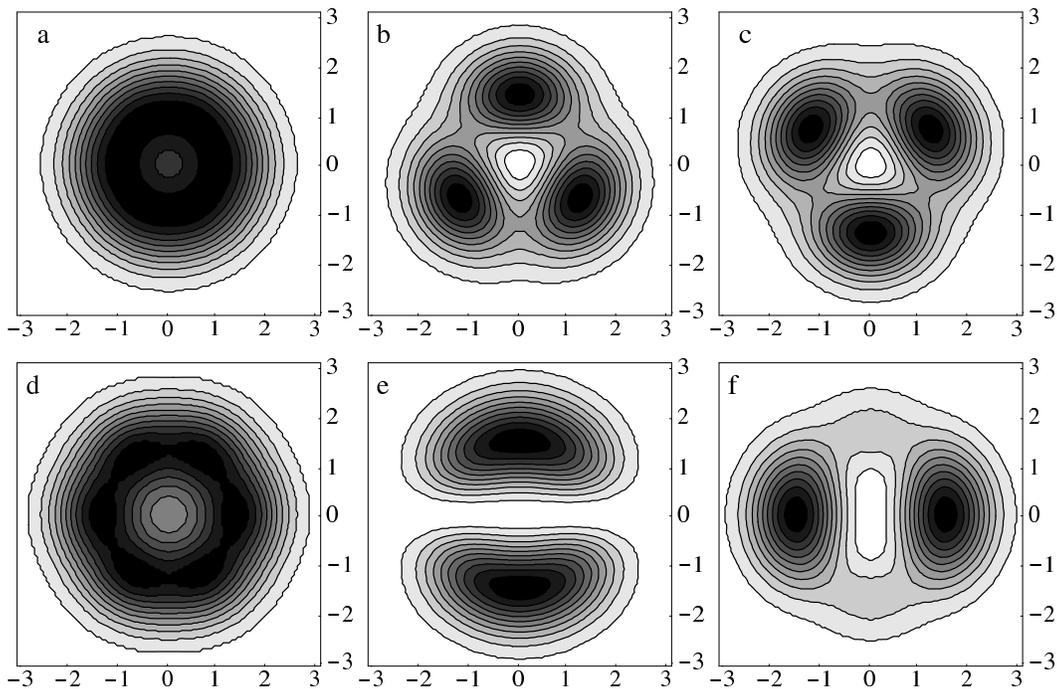, width=14cm}
\caption{UHF orbital \Ds\ $|\varphi_i|^2$ ($i=1,2,3$)
for $N=3$ and $S_z =3/2$. Upper row $\lambda=4$, lower row
$\lambda=6$. For the single-particle energies we obtain  
(a) $\varepsilon_1 = 4.92$ and (b), (c)  $\varepsilon_2 = \varepsilon_3 = 5.84$;
(d) $\varepsilon_1 = 6.44$ and (e), (f)  $\varepsilon_2 = \varepsilon_3 = 7.11$.}
\label{n3orbdich}
\end{figure*}
In \Abb{n3orbdich} we show the orbital \Ds\ for $\lambda=4$ and $\lambda=6$.
We find a sixfold orbital, as well as two diametrically oriented
threefold orbitals.
One clearly recognizes how the sixfold modulation 
results from this. Note that the HF orbitals are not localized
(for example at the angles of a triangle).

At this point we want to address a related issue,
the uniqueness of the HF orbitals.
One can easily show with the help of the HF equations  
that HF orbitals with the same spin are no longer unique, if
the corresponding one-particle energies $\varepsilon_i$ are degenerate.
In this case, any unitary transformation of
degenerate orbitals also fulfills the HF equations.
In \Abb{n3orbdich}, the energies  $\varepsilon_i$
are degenerate for the two states (b),(c) and (e),(f).
Therefore these two orbitals are no longer uniquely determined, -- 
in addition to the orientational degeneracy
of the total \SD\, which is physically obvious.

Now, we want to have a closer look on the orbital energies:
it is natural to presume that their degeneracies 
are a signature of Wigner crystallization, i.e.~the geometry
of the Wigner molecule.
For strong \IA\ one should be able to represent the system
as a lattice problem on an equilateral triangle.
The corresponding Hamiltonian for $N=3$, $S_z=3/2$  reads
\begin{equation}
\label{tuma3s3}
H_3 = \left( \begin{array}{rrr} U & -t & -t \\
				-t & U & -t \\
				-t & -t & U \end{array} \right) \; ,
\end{equation}
where $U$ is the on-site energy and $t$ is the tunneling matrix 
element between localized states. The eigenvalues of $H_3$ are
$\varepsilon_{1}=U-2t$ and twice $\varepsilon_{2/3} = U + t$
which is in fact the degeneracy of the UHF orbital energies
(\Abb{n3orbdich}).

On the other hand, for $S_z=1/2$ the tight-binding Hamiltonian involves 
tunneling only between the two spin up states and takes the form 
\begin{equation}
\label{tuma3s1}
H_3' = \left( \begin{array}{rrc} U & -t & 0 \\
				-t & U & 0 \\
			  	 0 & 0 & U \end{array} \right) \; .
\end{equation}
The eigenvalues are $\varepsilon_{1/2}=U \pm t$ (spin up) 
and $\varepsilon_3 = U $ (spin down). 
With UHF for $\lambda=6$ we find $\varepsilon_1 = 6.65$, 
$\varepsilon_2 = 7.10$ and $\varepsilon_3 = 6.87$, 
which has to be compared with the orbital energies
for the polarized state given in \Abb{n3orbdich} and yields $t\approx 0.22$. 
For larger $\lambda$ the agreement becomes  better, 
e.g.~for $\lambda=12$ we find  $\varepsilon_1 = 10.140$, 
$\varepsilon_2 = 10.309$
and $\varepsilon_3 = 10.224$ for $S_z=1/2$, while $\varepsilon_1 = 10.06$
and  $\varepsilon_{2/3} = 10.313$ for $S_z=3/2$, which gives $t\approx 0.084$
in both cases.

\subsection{Lattice Hamiltonian and localized orbitals}
\label{Laha}

For large $\lambda$ the HF Hamiltonian has the same eigenvalues 
as a lattice Hamiltonian.
Thus, there must be one-to-one correspondence
between these two.  
Remember, however, that HF is a one-particle picture and
thus the tight binding Hamiltonian describes {\it one} particle
hopping on a grid. 
The HF Hamiltonian is diagonal in the HF basis \refp{hforb},
\begin{equation}
 \label{hfham}
 \brkt{i}{h}{j}+\sum_k^N (i k|w|j k) =\varepsilon_i \delta_{ij} \;.
\end{equation} 
Now, if the eigenvalues $\varepsilon_i$ coincide with those 
of a lattice Hamiltonian, e.g.~$H_3$ in \refp{tuma3s3},
this means that we have to transform the UHF orbitals with the inverse
of the orthogonal transformation which diagonalizes the lattice Hamiltonian
to pass over to localized orbitals. The Slater determinant 
is not changed when we transform among occupied orbitals, \cite{edmin63}
\begin{equation}
 \label{locorb}
 \ket{p} = \sum_i^N o_p^i \ket{i} \;.
\end{equation} 
In this new basis the HF equations read
\begin{equation}
 \label{hfglgloc}
 \sum_q^N \big\{\brkt{p}{h}{q}+\sum_{r}^N (p r|w|q r)\big\} o^i_{q}
    = \varepsilon_i o^i_{p} \;.
\end{equation} 
Now, in the basis $\ket{p}$, we should have non vanishing $\brkt{p}{h}{q}$
only for nearest neighbors
\cite{foot4}
and the contribution of the two-particle matrix element should 
essentially be given by the direct term, i.e.~diagonal elements of the Coulomb
interaction. Then \refp{hfglgloc} reduces to
\begin{equation}
 \label{hftuma}
 \sum_q^N \big\{\brkt{p}{h}{q}+\delta_{p q}\sum_{r}^N (p r|w|p r) \big\} o^i_{q}
    = \varepsilon_i o^i_{p} \; ,
\end{equation}
which is now of the form of a lattice Hamiltonian.

We now present strong numerical evidence for 
this connection between the UHF Hamiltonian and a lattice Hamiltonian
for $N=4$ and 5 which are the simplest cases of electrons on a ring. 
For $N=4$, $S_z=2$ we have
\begin{equation}
\label{tuma4s2}
H_4 = \left( \begin{array}{rrrr} U & -t & 0  & -t \\
				-t & U  & -t & 0 \\
				 0 & -t & U  & -t \\
				-t & 0  & -t & U \end{array} \right) \; ,
\end{equation}
with the eigenvalues $\varepsilon_{1}=U-2t$, $\varepsilon_{2/3} = U$ and
$\varepsilon_{4}=U+2t$.
The eigenvectors of $H_4$ determine the transformation \refp{locorb}.
Applying this transformation to the HF Hamiltonian, as we did in
\refp{n2s1hfhatra}, we obtain for $\lambda = 8$ an Hamiltonian of the form
\refp{tuma4s2} with $U=10.924$  and $t=0.195$.
The next nearest neighbor hopping matrix element (hopping along the diagonal
of the square) is $t^*= 2\varepsilon_{2} - \varepsilon_{1}
-\varepsilon_{4}= 0.003$, which is indeed very small.

Likewise we can determine the lattice Hamiltonians for other \TZ s
and spin configurations and we have collected results for $t$ and $U$
for stronger interaction up to $\lambda = 20$.
For $N=4$, $S_z=1$ the lattice Hamiltonian reads
\begin{equation}
\label{tuma4s1}
H_4' = \left( \begin{array}{rrrc} U & -t & 0 & 0 \\
				-t & U & -t & 0 \\
				 0 & -t& U & 0 \\
				 0 & 0 &  0&\, U \end{array} \right) \; ,
\end{equation}
with the eigenvalues $\varepsilon_{1}=U-\sqrt 2 t$, $\varepsilon_{2} = U$ and
$\varepsilon_{3}=U+\sqrt 2 t$ (spin up) and $\varepsilon_{4} = U$ (spin down),
while for $N=4$, $S_z=0$ we have
\begin{equation}
\label{tuma4s0}
H_4'' = \left( \begin{array}{rrrr} U & -t & 0  & 0 \\
				-t & U  & 0  & 0 \\
				 0 & 0  & U  & -t \\
				 0 & 0  & -t & U \end{array} \right) \; ,
\end{equation}
with $\varepsilon_{1/2}=U \pm t$ (spin up), $\varepsilon_{3/4} = U
\pm t$ (spin down). Here, we have to assume that the four states
are occupied with two pairs of nearest neighbor parallel spins
in order to obtain agreement with the UHF orbital energies.
The values of $t$ we obtain in this way 
for the three spin states $S_z=0,1,2$ 
agree within 1\% for $\lambda = 8$ .

For $N=5$ we have a pentagon and again three different spin states.
For $S_z=5/2$ the lattice Hamiltonian with nearest neighbor hopping is
\begin{equation}
\label{tuma5s5}
H_5 = \left( \begin{array}{rrrrr} U & -t & 0  & 0  & -t \\
				 -t & U  & -t & 0  & 0 \\
				  0 & -t & U  & -t & 0 \\
				  0 & 0  & -t & U  & -t \\
				 -t & 0  & 0  & -t & U \end{array} \right) \; ,
\end{equation}
with the eigenvalues $\varepsilon_{1}=U-2t$, 
$\varepsilon_{2/3} = U + t(1-\sqrt 5)/2$ and 
$\varepsilon_{4/5}=U + t(1+\sqrt 5)/2$, while for $S_z=3/2$ we have
\begin{equation}
\label{tuma5s3}
H_5' = \left( \begin{array}{rrrrc} U & -t & 0  & 0  & 0 \\
				 -t & U  & -t & 0  & 0 \\
				  0 & -t & U  & -t & 0 \\
				  0 & 0  & -t & U  & 0  \\
				  0 & 0  & 0  & 0  & \, U \end{array} \right) \; ,
\end{equation}
with $\varepsilon_{1/2} = U - t(\sqrt 5 \pm 1)/2 $,
$\varepsilon_{3/4}=U + t(\sqrt 5 \mp 1)/2$ (spin up) and
$\varepsilon_{5}=U$ (spin down). 
Finally for $S_z=1/2$ we have
\begin{equation}
\label{tuma5s1}
H_5'' = \left( \begin{array}{rrrcc} U & 0 & 0  & 0  & 0 \\
				  0 & U  & -t & 0  & 0 \\
				  0 & -t & U  & 0  & 0 \\
				  0 & 0  & 0  & \, U  & 0  \\
				  0 & 0  & 0  & 0  & \, U \end{array} \right) \; ,
\end{equation}
with the eigenvalues $\varepsilon_{1/3} = U \pm t$,
$\varepsilon_{2}=U$ (spin up) and $\varepsilon_{4/5}=U$ (spin down).
Note that here the values of the UHF orbital energies suggest a model with
only two nearest neighbor parallel spins.
For $\lambda = 6$ the values of $t$ for all three spin states
coincide within 1\%.

\begin{figure}
\centering\epsfig{file=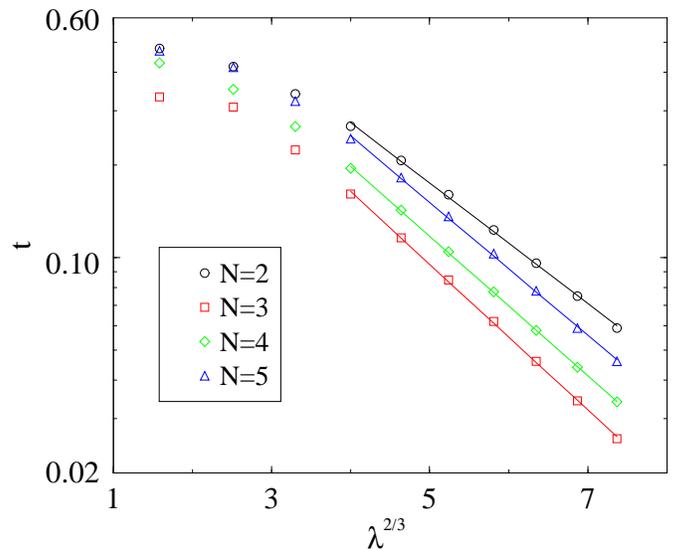, width=9cm}
\caption{Log-linear plot of tunnel matrix element $t$ vs. $ \lambda^{2/3}$
for various \TZ s. For $\lambda \ge 8$ the line of best fit is shown.}
\label{tvslambda}
\end{figure}
Figure \ref{tvslambda} summarizes our findings about the tunnel matrix elements. 
Reference \onlinecite{haeus00} predicts $t \propto \exp (-\sqrt{r_s})$, where
$r_s$ is the nearest neighbor distance
of the electrons measured in units of the effective Bohr radius.
Since classically $r_s \propto \lambda^{4/3}$ (cf.~Appendix \ref{Klako}) 
we plot $\ln t$ versus $ \lambda^{2/3}$. 
For $\lambda \agt 8$ we find indeed a linear behavior.
For lower $\lambda$, the tunneling matrix element is not really defined, 
since the lattice model is not appropriate.
The tunneling matrix element is largest for $N=2$
because two electrons are always closest (see Appendix \ref{Klako}).
Three electrons always have the smallest value of $t$
because the corresponding equilateral triangle has a longer side 
than the square and the pentagon.
For higher \TZ s one electron enters the center of the dot,
and the UHF spectra are more complicate but still show
the typical degeneracies.
However, now the lattice Hamiltonian has various tunneling constants 
and on-site energies.

\subsection{Seven-\E\ Wigner molecule} 
\label{HFn7}

\begin{figure}
\psfrag{a}{(a)}
\psfrag{b}{(b)}
\psfrag{c}{(c)}
\psfrag{d}{(d)}
\psfrag{x}{{\large $\tilde{r}_s$}}
\epsfig{file=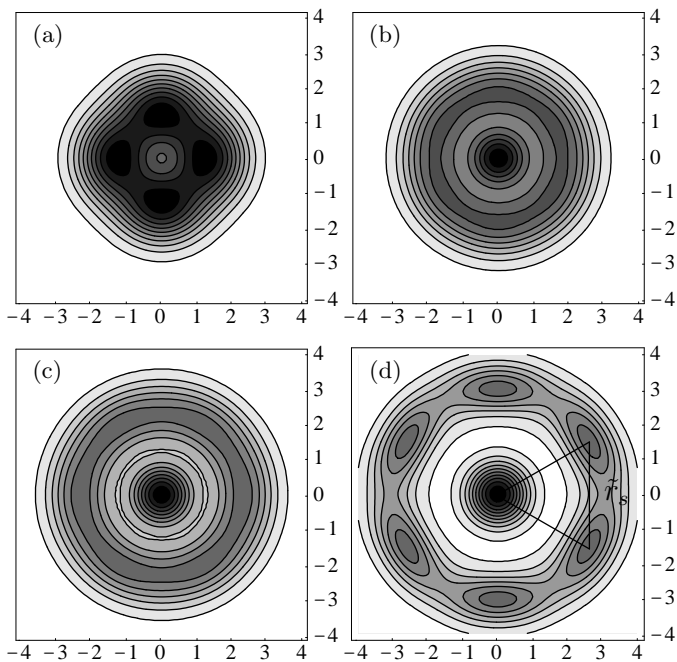, width=9cm}
\caption{One-particle densities for the UHF \Grdz\
of  $N=7$ \E s. (a) $\lambda=1$, (b) $\lambda=2$, both $S_z=1/2$.
(c) $\lambda=4$, (d) $\lambda=10$, both $S_z=7/2$.}
\label{n7dich}
\end{figure}
Seven classical \E s form a equilateral hexagon with one 
central \E , which is a fragment of a hexagonal lattice.
In \Abb{n7dich} we show UHF \Ds\ for $N=7$ starting
with a small $\lambda$.
The UHF \Grdz\ is  $S_z=1/2$ up to $\lambda\alt 3$, 
then spin-polarized. In  \Abb{n7dich}(a) for $\lambda = 1$ we see a fourfold
modulated \D . How is that possible for seven \E s?
The answer is that in this case the energy shell picture of the
harmonic oscillator is still valid: six \E s are just a shell closure
and the next electron is put in the new shell in an orbital with maximal
angular momentum. This angular momentum is
$M=\pm 2$ and from the superposition one obtains a fourfold standing wave
(cf.~Ref.~\onlinecite{hiros99}).
Here, the \En\ is basically the same as in RHF, but the \SD\ breaks
the symmetry. 

With increasing interaction strength a Wigner molecule is formed
with one electron in the center and six in the surrounding ring 
[\Abb{n7dich}(b)-(c)].
We want to emphasize that the UHF \Ds\ mirror the classical
shell filling. 
This can even be quantified: the positions of the maxima 
(even in the 'round' densities) agree very well with the classical
configurations in Appendix \ref{Klako}. From the UHF density the 
nearest neighbor distance $\tilde{r}_s$  can be determined.
For example from \Abb{n7dich}(d) we find $\tilde{r}_s\approx 3.0$,
which is also the classical value. 
Here we have to take into account that we measure length in
oscillator units. Frequently, one is interested in the density
parameter $r_s$ given in effective Bohr radii.
\cite{foot5}
Then \Abb{n7dich}(d) gives
$r_s = \tilde{r}_s l_0/a^*_{\Be}= \lambda \tilde{r}_s \approx 30$.
The $r_s$ values we obtain in this way agree also well
with the results of Ref.~\onlinecite{egger99}. 
There $r_s$ is determined from the 
first maximum of the two-particle correlation function.

\section{Unrestricted Hartree-Fock with a magnetic field}
\label{Maguhf}

In this section we want to present some calculations
with a magnetic field orthogonal to the plane of the
\QP . This system has been discussed extensively in the 
literature, especially in connection with the quantum
Hall effect. UHF calculations by M\"uller and Koonin \cite{muell96} 
have shown a {\it magnetic field induced Wigner
crystallization}. However, they only considered the limiting
case of a strong magnetic field and therefore included in the basis
for expanding the UHF orbitals only states from the lowest Landau level
(Fock-Darwin levels with $n=0$). The high field case has also been studied
by Palacios et al.\cite{palac94} and Ruan et al.\cite{chan01,ruan00,ruan99} 
To study smaller magnetic fields, our basis is better adjusted to the problem.
It is intuitively clear, that \E s are further localized by
the magnetic field. Indeed, for sufficiently strong fields,
we do not find an even-odd effect
for UHF \Ds\ but molecule-like \Ds\ for all \TZ s. 

Numerically, thanks to the similar form of the Hamiltonian
\refp{hamag2} to the one without magnetic field, the generalization of
our UHF code is straightforward. 
However, the magnetic field breaks time reversal symmetry,
left and right turning solutions are no longer energetically degenerate.
Therefore in the expansion of the UHF orbitals \refp{hforb} we have to use
complex coefficients. 

We first consider three \E s and a large interaction parameter
$\lambda = 10$. This means that we have a shallow \QP\
where the Coulomb \IA\ dominates and the magnetic field is relatively weak.
In \Abb{n3wcdich} we display the evolution 
of the UHF one-particle densities with increasing magnetic field strength
 $\tilde \omega_c = \omega_c/\omega$ at fixed $\lambda$.
This is not exactly the physical situation, corresponding to a
\QP\ exposed to an  increasing magnetic field, since
the \CC\ $\lambda$ becomes smaller with increasing field.
Here we just want to show that a magnetic field does not have the
same effect on the UHF \D\ as a strong \IA.
\begin{figure}
\psfrag{a}{(a)}
\psfrag{b}{(b)}
\psfrag{c}{(c)}
\psfrag{d}{(d)}
\epsfig{file=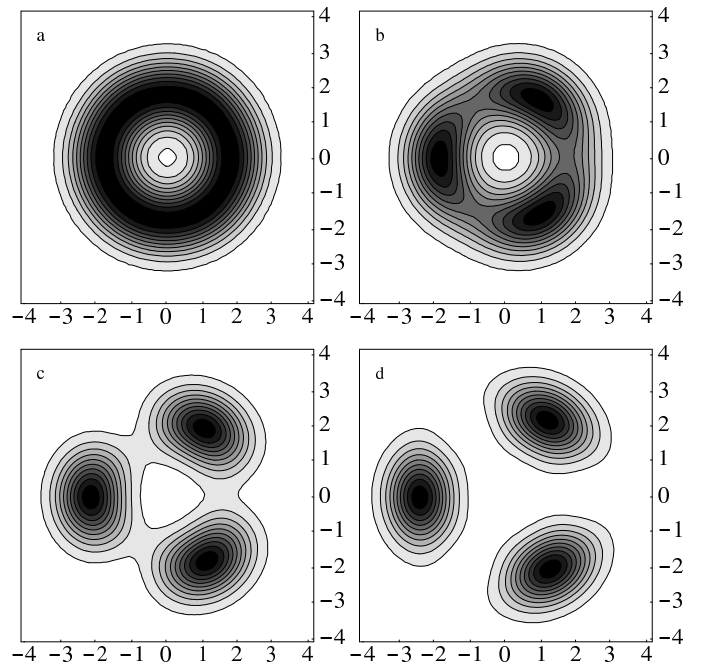, width=9cm}
\caption{Evolution of the UHF one-particle densities for  $N=3$, $S_z=3/2$ 
and $\lambda=10$ with increasing magnetic field strength
$\tilde \omega_c = \omega_c/\omega$. (a) $\tilde \omega_c=0$, 
(b) $\tilde \omega_c=0.5$, (c) $\tilde \omega_c=1.5 $, 
(d) $\tilde \omega_c=2.5$.}
\label{n3wcdich}
\end{figure}

In \Abb{n3wcdich}(d) we see three distinct, localized \E s
in the UHF \D . The three single orbital densities have nearly the same 
form. They are thus similar to the orbitals chosen in 
Ref.~\onlinecite{szafr03}.  
With {\it decreasing} magnetic field strength
the maxima in azimuthal direction vanish slowly, until
we have again a nearly round density for $\omega_c=0$ as in \Abb{evenodd}(a).
The \D\ in \Abb{n3wcdich}(a) has been obtained from an initial guess
with threefold symmetry. Therefore we can be sure that we have not obtained
a local minimum but the true HF \Grdz .

\begin{figure*}
\psfrag{a}{(a)}
\psfrag{b}{(b)}
\psfrag{c}{(c)}
\psfrag{d}{(d)}
\psfrag{e}{(e)}
\psfrag{f}{(f)}
\centering \epsfig{file=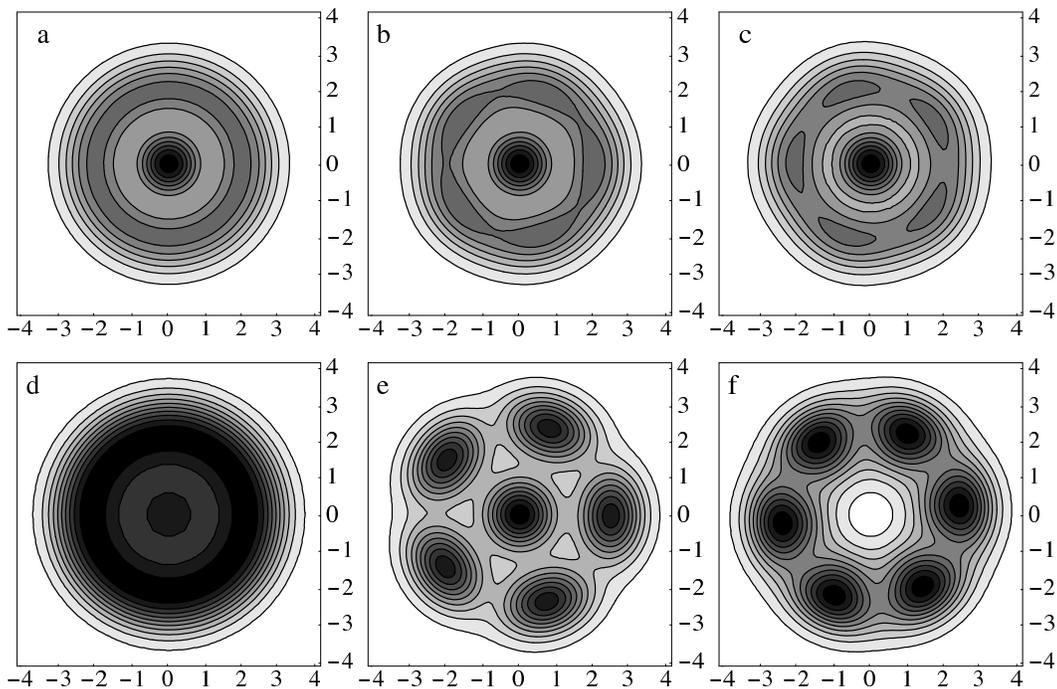, width=14cm}
\caption{Evolution of the UHF one-particle \D\ for $N=6$, 
$S_z=3$ and $\lambda=3.2$ with increasing magnetic field strength, 
(a) $\tilde \omega_c = 0$, (b) $\tilde \omega_c = 0.1$, 
(c) $\tilde \omega_c = 0.5 $, (d) $\tilde \omega_c = 1$, 
(e) $\tilde \omega_c= 2 $, (f) $\tilde \omega_c = 2.5 $. In (f) sixfold
isomer with energy $E^*_{\HF} = 45.182$.}
\label{n6wcdich}
\end{figure*}

As a second example we show the evolution of the UHF \D\
of six \E s at intermediate coupling strength $\lambda=3.2$. 
Without magnetic field the \D\ is round, \Abb{n6wcdich}(a), and with 
a weak magnetic field fivefold with a central \E , Figs. \ref{n6wcdich}(b),(c).
Remarkably, 
for intermediate magnetic field $\tilde \omega_c \approx 1 \ldots 1.5$,
the UHF \Grdz\ has a perfectly round \D , \Abb{n6wcdich}(d), and also
a rotationally symmetric \SD.
This is the so-called maximum-density-droplet of MacDonald 
et al.,\cite{macdo93}
where the \E s occupy the lowest orbitals with increasing angular momentum.
Here the orbitals with $M=0$,1,2,3,4,5 are occupied, and the UHF
solution is identical to the RHF solution with total angular momentum
$M^{\ges}=15$. 

Finally, in \Abb{n6wcdich}(e) for strong magnetic field 
we have a distinctly localized fivefold Wigner molecule. 
Figure \ref{n6wcdich}(f) for $\tilde \omega_c = 2.5$ 
shows a sixfold isomer which is higher in energy by 0.009 than the 
fivefold \Grdz .

\section{Conclusion}
\label{conclusion}
In conclusion, we have discussed the properties of unrestricted
Hartree-Fock (UHF) calculations for electrons in a quantum dot, focusing
on the regime of strong correlations, when the electrons begin to
form a Wigner molecule. The UHF energies are good estimates of the true 
ground-state energies, especially for the polarized states, even at strong 
\IA . In this regime, the UHF energies become nearly spin independent,
faster than it is the case for the true energies.
However, the energy differences between different spin states cannot
be resolved correctly by UHF, the polarized state is unphysically favored for
stronger \IA . 

Regarding the interpretation of other quantities obtained 
from the UHF Slater determinant, we have shown that considerable caution must
be taken: we find deformed densities in the regime of intermediate \IA\ 
$\lambda \approx 1 \ldots 4$.
For stronger interaction the densities are azimuthal\-ly modulated 
for an even number of electrons per spatial shell, and round for an
odd number per shell. The onset of this modulation is enhanced 
within UHF, so that UHF leads to an overestimation of the value of 
the critical density for the crossover to the Wigner molecule.
We want to emphasize that the even-odd effect we found
is an artifact of the symmetry breaking of UHF and arises
from a degeneracy of states with opposite total angular momentum.

For very strong \IA , we have shown that the UHF Hamiltonian
corresponds to a tight-binding model of a particle hopping between
the sites of the Wigner molecule. From the UHF orbital energies
we have obtained the hopping matrix elements. This correspondence
explains why the UHF energies become nearly spin independent which is 
expected for localized electrons and was not found with restricted HF.

The maxima of the UHF densities mirror the classical
filling scheme with the electrons arranged in spatial shells. 
In contrast, the UHF two particle density (conditional probability
density) has no direct physical meaning, because the UHF method can
not take correlations  properly into account.
Finally, in a strong magnetic field the UHF densities are
always  molecule-like and there is no even-odd effect.

The numerical complexity of the UHF method is comparable
to the frequently used density-functional approach.
However, as shown here, UHF has the advantage to cope also with the
strongly interacting limit and gives further physical insight
in that case. 
For the tiny energy differences which determine the spin ordering
or the addition energies at $\lambda \agt 2$ one has to employ the
computationally more expensive quantum Monte Carlo methods.

\begin{acknowledgments}
We acknowledge useful discussions with Alessandro De Martino, 
Wolfgang H\"ausler, Christoph Theis and Till Vorrath.
This work has been supported by the Deutsche Forschungsgemeinschaft
(SFB 276).
\end{acknowledgments}

\begin{appendix}
\section{Configurations of classical point charges}
\label{Klako}

\begin{table}[b]
\label{klasskonf}
\begin{ruledtabular}
\renewcommand{\arraystretch}{1.8}
\vspace*{1cm}
\begin{tabular}{ccccc} 
 $N$ & Geometry & $r_a^3/\lambda$ & $r_s/\lambda^{4/3}$ &
$E/r_a^2$\\ \hline
$2$ & dumbbell (2) & $\frac{1}{4}$ & $ \approx 1.260$ & $3$\\ 

 $3$ & triangle (3) & $\frac{1}{\sqrt{3}}\approx 0.577$  & $ \approx 1.442 $  & 
 $\frac{9}{2}$\\

 $4$ & square (4) & $\frac{1}{4}+\frac{1}{\sqrt{2}} \approx 0.957 $  &
 $\approx 1.394 $  & $6$\\

 $5$ & pentagon (5) & $\sqrt{1+\frac{2}{\sqrt{5}}} \approx 1.376 $ &
 $\approx 1.308 $ & $\frac{15}{2}$\\

 $5^*$ & square  (4,1)& $\frac{5}{4}+\frac{1}{\sqrt{2}} \approx 1.957 $ &
 $ \approx 1.251 $  & $6$\\

 $6$ & pentagon (5,1)& $1+\sqrt{1+\frac{2}{\sqrt{5}}} $ &
 $ \approx 1.334 $  & $\frac{15}{2}$\\

 $6^*$ & hexagon (6) & $\frac{5}{4}+\frac{1}{\sqrt{3}} \approx 1.827 $ &
 $\approx 1.223 $  & $9$\\

 $7$ & hexagon (6,1) &  $\frac{9}{4}+\frac{1}{\sqrt{3}} \approx 2.827 $ &
 $\approx 1.414 $  & $9$\\
\end{tabular}
\end{ruledtabular}
\end{table}
In Table \ref{klasskonf} we give the classical configurations
for up to seven 2D electrons in a parabolic confinement potential
with zero magnetic field.
$r_a$ is the distance of the outer electrons from  the center
measured in oscillator length $l_0$.
$r_s$ is the nearest neighbor distance measured in effective Bohr radii
$a^*_{\Be}$. 
Energies are given in units of $\hbar\omega$. These quantities depend only 
on $N$ and $\lambda$. 

For $N=5$ and 6 we specify isomers with higher energies.
Due to the classical virial theorem there is a simple
relationship between the energy and $r_a$. When we denote
the distance of the $i$-th electron from the center by $r_i$,
we have
\begin{equation}
\label{eklass}
E = \frac{3}{2} \: \sum_{i=1}^N r_i^2 \; .
\end{equation}

\end{appendix}


\begin{thebibliography}{31}
\expandafter\ifx\csname natexlab\endcsname\relax\def\natexlab#1{#1}\fi
\expandafter\ifx\csname bibnamefont\endcsname\relax
  \def\bibnamefont#1{#1}\fi
\expandafter\ifx\csname bibfnamefont\endcsname\relax
  \def\bibfnamefont#1{#1}\fi
\expandafter\ifx\csname citenamefont\endcsname\relax
  \def\citenamefont#1{#1}\fi
\expandafter\ifx\csname url\endcsname\relax
  \def\url#1{\texttt{#1}}\fi
\expandafter\ifx\csname urlprefix\endcsname\relax\def\urlprefix{URL }\fi
\providecommand{\bibinfo}[2]{#2}
\providecommand{\eprint}[2][]{\url{#2}}

\bibitem[{\citenamefont{Pfannkuche et~al.}(1993)\citenamefont{Pfannkuche,
  Gudmundsson, and Maksym}}]{pfann93}
\bibinfo{author}{\bibfnamefont{D.}~\bibnamefont{Pfannkuche}},
  \bibinfo{author}{\bibfnamefont{V.}~\bibnamefont{Gudmundsson}},
  \bibnamefont{and} \bibinfo{author}{\bibfnamefont{P.~A.}
  \bibnamefont{Maksym}}, \bibinfo{journal}{Phys. Rev. B}
  \textbf{\bibinfo{volume}{47}}, \bibinfo{pages}{2244} (\bibinfo{year}{1993}).

\bibitem[{\citenamefont{Palacios et~al.}(1994)\citenamefont{Palacios,
  Mart{\'i}n-Moreno, Chiappe, Louis, and Tejedor}}]{palac94}
\bibinfo{author}{\bibfnamefont{J.~J.} \bibnamefont{Palacios}},
  \bibinfo{author}{\bibfnamefont{L.}~\bibnamefont{Mart{\'i}n-Moreno}},
  \bibinfo{author}{\bibfnamefont{G.}~\bibnamefont{Chiappe}},
  \bibinfo{author}{\bibfnamefont{E.}~\bibnamefont{Louis}}, \bibnamefont{and}
  \bibinfo{author}{\bibfnamefont{C.}~\bibnamefont{Tejedor}},
  \bibinfo{journal}{Phys. Rev. B} \textbf{\bibinfo{volume}{50}},
  \bibinfo{pages}{5760} (\bibinfo{year}{1994}).

\bibitem[{\citenamefont{Fujito et~al.}(1996)\citenamefont{Fujito, Natori, and
  Yasunaga}}]{fujit96}
\bibinfo{author}{\bibfnamefont{M.}~\bibnamefont{Fujito}},
  \bibinfo{author}{\bibfnamefont{A.}~\bibnamefont{Natori}}, \bibnamefont{and}
  \bibinfo{author}{\bibfnamefont{H.}~\bibnamefont{Yasunaga}},
  \bibinfo{journal}{Phys. Rev. B} \textbf{\bibinfo{volume}{53}},
  \bibinfo{pages}{9952} (\bibinfo{year}{1996}).

\bibitem[{\citenamefont{M{\"u}ller and Koonin}(1996)}]{muell96}
\bibinfo{author}{\bibfnamefont{H.-M.} \bibnamefont{M{\"u}ller}}
  \bibnamefont{and} \bibinfo{author}{\bibfnamefont{S.~E.}
  \bibnamefont{Koonin}}, \bibinfo{journal}{Phys. Rev. B}
  \textbf{\bibinfo{volume}{54}}, \bibinfo{pages}{14532} (\bibinfo{year}{1996}).

\bibitem[{\citenamefont{Yannouleas and Landman}(1999)}]{yanno}
\bibinfo{author}{\bibfnamefont{C.}~\bibnamefont{Yannouleas}} \bibnamefont{and}
  \bibinfo{author}{\bibfnamefont{U.}~\bibnamefont{Landman}},
  \bibinfo{journal}{Phys. Rev. Lett.} \textbf{\bibinfo{volume}{82}},
  \bibinfo{pages}{5325} (\bibinfo{year}{1999}); \bibinfo{note}{{\bf 85},
2220(E) (2000); Phys.~Rev.~B
  {\bf 61}, 15895 (1999); J.~Phys.~Condens.~Matter {\bf 14}, L591 (2000)}.

\bibitem[{\citenamefont{Reusch et~al.}(2001)\citenamefont{Reusch, H{\"a}usler,
  and Grabert}}]{reusc01}
\bibinfo{author}{\bibfnamefont{B.}~\bibnamefont{Reusch}},
  \bibinfo{author}{\bibfnamefont{W.}~\bibnamefont{H{\"a}usler}},
  \bibnamefont{and} \bibinfo{author}{\bibfnamefont{H.}~\bibnamefont{Grabert}},
  \bibinfo{journal}{Phys. Rev. B} \textbf{\bibinfo{volume}{63}},
  \bibinfo{pages}{113313} (\bibinfo{year}{2001}).

\bibitem[{\citenamefont{Sundqvist et~al.}(2002)\citenamefont{Sundqvist,
  Volokov, Lozovik, and Willander}}]{sundq02}
\bibinfo{author}{\bibfnamefont{P.~A.} \bibnamefont{Sundqvist}},
  \bibinfo{author}{\bibfnamefont{S.~Y.} \bibnamefont{Volokov}},
  \bibinfo{author}{\bibfnamefont{Y.~E.} \bibnamefont{Lozovik}},
  \bibnamefont{and}
  \bibinfo{author}{\bibfnamefont{M.}~\bibnamefont{Willander}},
  \bibinfo{journal}{Phys. Rev. B} \textbf{\bibinfo{volume}{66}},
  \bibinfo{pages}{075335} (\bibinfo{year}{2002}).

\bibitem[{\citenamefont{Szafran et~al.}(2003)\citenamefont{Szafran, Bednarek,
  and Adamowski}}]{szafr03}
\bibinfo{author}{\bibfnamefont{B.}~\bibnamefont{Szafran}},
  \bibinfo{author}{\bibfnamefont{S.}~\bibnamefont{Bednarek}}, \bibnamefont{and}
  \bibinfo{author}{\bibfnamefont{J.}~\bibnamefont{Adamowski}},
  \bibinfo{journal}{Phys. Rev. B} \textbf{\bibinfo{volume}{67}},
  \bibinfo{pages}{045311} (\bibinfo{year}{2003}).

\bibitem[{\citenamefont{Reimann and Manninen}(2002)}]{reima02}
\bibinfo{author}{\bibfnamefont{S.~M.} \bibnamefont{Reimann}} \bibnamefont{and}
  \bibinfo{author}{\bibfnamefont{M.}~\bibnamefont{Manninen}},
  \bibinfo{journal}{Rev. Mod. Phys.} \textbf{\bibinfo{volume}{74}},
  \bibinfo{pages}{1283} (\bibinfo{year}{2002}).

\bibitem[{\citenamefont{Tarucha et~al.}(1996)\citenamefont{Tarucha, Austing,
  Honda, van~der Hage, and Kouwenhoven}}]{taruc96}
\bibinfo{author}{\bibfnamefont{S.}~\bibnamefont{Tarucha}},
  \bibinfo{author}{\bibfnamefont{D.~G.} \bibnamefont{Austing}},
  \bibinfo{author}{\bibfnamefont{T.}~\bibnamefont{Honda}},
  \bibinfo{author}{\bibfnamefont{R.~J.} \bibnamefont{van~der Hage}},
  \bibnamefont{and} \bibinfo{author}{\bibfnamefont{L.~P.}
  \bibnamefont{Kouwenhoven}}, \bibinfo{journal}{Phys. Rev. Lett.}
  \textbf{\bibinfo{volume}{77}}, \bibinfo{pages}{3613} (\bibinfo{year}{1996}).

\bibitem[{\citenamefont{Wigner}(1938)}]{wigne38}
\bibinfo{author}{\bibfnamefont{E.~P.} \bibnamefont{Wigner}},
  \bibinfo{journal}{Trans. Faraday Soc.} \textbf{\bibinfo{volume}{34}},
  \bibinfo{pages}{678} (\bibinfo{year}{1938}).

\bibitem[{\citenamefont{Zhitenev et~al.}(1999)\citenamefont{Zhitenev, Brodsky,
  Ashoori, Pfeiffer, and West}}]{zhite99}
\bibinfo{author}{\bibfnamefont{N.~B.} \bibnamefont{Zhitenev}},
  \bibinfo{author}{\bibfnamefont{M.}~\bibnamefont{Brodsky}},
  \bibinfo{author}{\bibfnamefont{R.~C.} \bibnamefont{Ashoori}},
  \bibinfo{author}{\bibfnamefont{L.~N.} \bibnamefont{Pfeiffer}},
  \bibnamefont{and} \bibinfo{author}{\bibfnamefont{K.~W.} \bibnamefont{West}},
  \bibinfo{journal}{Science} \textbf{\bibinfo{volume}{285}},
  \bibinfo{pages}{715} (\bibinfo{year}{1999}).

\bibitem[{\citenamefont{Reimann et~al.}(2000)\citenamefont{Reimann, Koskinen,
  and Manninen}}]{reima00}
\bibinfo{author}{\bibfnamefont{S.~M.} \bibnamefont{Reimann}},
  \bibinfo{author}{\bibfnamefont{M.}~\bibnamefont{Koskinen}}, \bibnamefont{and}
  \bibinfo{author}{\bibfnamefont{M.}~\bibnamefont{Manninen}},
  \bibinfo{journal}{Phys. Rev. B} \textbf{\bibinfo{volume}{62}},
  \bibinfo{pages}{8108} (\bibinfo{year}{2000}).

\bibitem[{\citenamefont{Hirose and Wingreen}(1999)}]{hiros99}
\bibinfo{author}{\bibfnamefont{K.}~\bibnamefont{Hirose}} \bibnamefont{and}
  \bibinfo{author}{\bibfnamefont{N.~S.} \bibnamefont{Wingreen}},
  \bibinfo{journal}{Phys.~Rev.~B} \textbf{\bibinfo{volume}{59}},
  \bibinfo{pages}{4604} (\bibinfo{year}{1999}).

\bibitem[{\citenamefont{Ruan et~al.}(1995)\citenamefont{Ruan, Liu, Bao, and
  Zhang}}]{ruan95}
\bibinfo{author}{\bibfnamefont{W.~Y.} \bibnamefont{Ruan}},
  \bibinfo{author}{\bibfnamefont{Y.~Y.} \bibnamefont{Liu}},
  \bibinfo{author}{\bibfnamefont{C.~G.} \bibnamefont{Bao}}, \bibnamefont{and}
  \bibinfo{author}{\bibfnamefont{Z.~Q.} \bibnamefont{Zhang}},
  \bibinfo{journal}{Phys. Rev. B} \textbf{\bibinfo{volume}{51}},
  \bibinfo{pages}{7942} (\bibinfo{year}{1995}).

\bibitem[{\citenamefont{H{\"a}usler}(2000)}]{haeus00}
\bibinfo{author}{\bibfnamefont{W.}~\bibnamefont{H{\"a}usler}},
  \bibinfo{journal}{Europhys. Lett.} \textbf{\bibinfo{volume}{49}},
  \bibinfo{pages}{231} (\bibinfo{year}{2000}).

\bibitem[{\citenamefont{Egger et~al.}(1999)\citenamefont{Egger, H{\"a}usler,
  Mak, and Grabert}}]{egger99}
\bibinfo{author}{\bibfnamefont{R.}~\bibnamefont{Egger}},
  \bibinfo{author}{\bibfnamefont{W.}~\bibnamefont{H{\"a}usler}},
  \bibinfo{author}{\bibfnamefont{C.~H.} \bibnamefont{Mak}}, \bibnamefont{and}
  \bibinfo{author}{\bibfnamefont{H.}~\bibnamefont{Grabert}},
  \bibinfo{journal}{Phys. Rev. Lett.} \textbf{\bibinfo{volume}{82}},
  \bibinfo{pages}{3320} (\bibinfo{year}{1999}), \bibinfo{note}{{\bf 83}, 462(E)
  1999}.

\bibitem[{\citenamefont{Chan and Ruan}(2001)}]{chan01}
\bibinfo{author}{\bibfnamefont{K.~S.} \bibnamefont{Chan}} \bibnamefont{and}
  \bibinfo{author}{\bibfnamefont{W.~Y.} \bibnamefont{Ruan}},
  \bibinfo{journal}{J.~Phys.~Condens.~Matter}
  \textbf{\bibinfo{volume}{13}}, \bibinfo{pages}{5799} (\bibinfo{year}{2001}).

\bibitem[{\citenamefont{Ruan et~al.}(2000)\citenamefont{Ruan, Chan, Ho, and
  Pun}}]{ruan00}
\bibinfo{author}{\bibfnamefont{W.~Y.} \bibnamefont{Ruan}},
  \bibinfo{author}{\bibfnamefont{K.~S.} \bibnamefont{Chan}},
  \bibinfo{author}{\bibfnamefont{H.~P.} \bibnamefont{Ho}}, \bibnamefont{and}
  \bibinfo{author}{\bibfnamefont{E.~Y.~B.} \bibnamefont{Pun}},
  \bibinfo{journal}{J.~Phys.~Condens.~Matter}
  \textbf{\bibinfo{volume}{12}}, \bibinfo{pages}{3911} (\bibinfo{year}{2000}).

\bibitem[{\citenamefont{Ruan and Cheung}(1999)}]{ruan99}
\bibinfo{author}{\bibfnamefont{W.~Y.} \bibnamefont{Ruan}} \bibnamefont{and}
  \bibinfo{author}{\bibfnamefont{H.-F.} \bibnamefont{Cheung}},
  \bibinfo{journal}{J.~Phys.~Condens.~Matter}
  \textbf{\bibinfo{volume}{11}}, \bibinfo{pages}{435} (\bibinfo{year}{1999}).

\bibitem[{\citenamefont{Mikhailov}(2002)}]{mikha02}
\bibinfo{author}{\bibfnamefont{S.~A.} \bibnamefont{Mikhailov}},
  \bibinfo{journal}{Phys. Rev. B} \textbf{\bibinfo{volume}{65}},
  \bibinfo{pages}{115312} (\bibinfo{year}{2002}).

\bibitem[{\citenamefont{Taut}(1993)}]{taut93}
\bibinfo{author}{\bibfnamefont{M.}~\bibnamefont{Taut}}, \bibinfo{journal}{Phys.
  Rev. A} \textbf{\bibinfo{volume}{48}}, \bibinfo{pages}{3561}
  (\bibinfo{year}{1993}).

\bibitem[{\citenamefont{MacDonald et~al.}(1993)\citenamefont{MacDonald, Yang,
  and Johnson}}]{macdo93}
\bibinfo{author}{\bibfnamefont{A.~H.} \bibnamefont{MacDonald}},
  \bibinfo{author}{\bibfnamefont{S.~E.} \bibnamefont{Yang}}, \bibnamefont{and}
  \bibinfo{author}{\bibfnamefont{M.~D.} \bibnamefont{Johnson}},
  \bibinfo{journal}{Aust. J. Phys.} \textbf{\bibinfo{volume}{46}},
  \bibinfo{pages}{345} (\bibinfo{year}{1993}).

\bibitem{foot6}{Note that $\Omega_c \leq 2$. In other words,
for a given material with effective Bohr radius $a^*_{\Be}$ and
given magnetic field $\omega_c$ there is a maximal \CC\
$\lambda_m := \sqrt{2} l_m/a^*_{\Be}$, where $l_m=\sqrt{\hbar/m\omega_c}$.
In particular, $\lambda_m \to 0$ for $\omega_c \to \infty$.}

\bibitem[{\citenamefont{Fock}(1930)}]{fock30}
\bibinfo{author}{\bibfnamefont{V.}~\bibnamefont{Fock}}, \bibinfo{journal}{Z.
  Phys.} \textbf{\bibinfo{volume}{61}}, \bibinfo{pages}{126}
  (\bibinfo{year}{1930}).

\bibitem{foot1}{The Hamiltonian $H$ and the total angular momentum commute, 
therefore the exact (non degenerate) ground state
must be an eigenstate of $L_z^{\ges}$. On the other hand, 
if we calculate the expectation value for the UHF ground state
one sees that the result is not necessarily integer.
However, only eigenstates of the total angular momentum 
are rotationally invariant.}

\bibitem{foot2}{An orbital with angular momentum $M\! =\! 1$ has
an isotropic density. Superposition of $M=\pm 1$ orbitals gives a
dumbbell formed density.}

\bibitem{foot2a}{Here we disagree with the authors of
Ref.~\onlinecite{yanno} who state that the degree of Wigner crystallization
can be extracted from the UHF CPD.}

\bibitem{foot3}{In Ref.~\onlinecite{muell96} the \IA\ constant was
$\lambda\approx 1.9$.}

\bibitem[{\citenamefont{Edmindston and Ruedenberg}(1963)}]{edmin63}
\bibinfo{author}{\bibfnamefont{C.}~\bibnamefont{Edmindston}} \bibnamefont{and}
  \bibinfo{author}{\bibfnamefont{K.}~\bibnamefont{Ruedenberg}},
  \bibinfo{journal}{Rev. Mod. Phys.} \textbf{\bibinfo{volume}{35}},
  \bibinfo{pages}{457} (\bibinfo{year}{1963}).

\bibitem{foot4}{Note that in Eq.~\refp{locorb} we transform only among
occupied orbitals with the same spin, $o^i_p \propto \delta_{\sigma_p \sigma_i}$
 and thus $\brkt{p}{h}{q}\propto \delta_{\sigma_p \sigma_q}$.}

\bibitem{foot5}{We note that various definitions of the
so-called Brueckner parameter $r_s$ are used in the literature.
Originally $r_s$ was defined
for homogeneous systems via the density $n_0 = 1/(\pi r_s^2)$.
The value corresponds to the area occupied by each \E\
and is roughly half of our $r_s$. This has to be taken into consideration
when comparing results by different authors.}
\end{thebibliography}

\end{document}